\documentclass[11pt]{article}
\usepackage{latexsym,amsmath,amssymb,multirow,bbm,nicefrac}
\usepackage{epstopdf,indentfirst,xcolor}
\definecolor{MyBlue}{RGB}{9, 31, 146}
\usepackage[colorlinks=true,urlcolor=MyBlue,citecolor=red,linkcolor=red]{hyperref}
\usepackage{enumitem}
\renewcommand{\a}{{\alpha}}
\renewcommand{\b}{{\beta}}

\newcommand{\dl}{{\delta}}
\newcommand{\eps}{{\epsilon}}
\newcommand{\pa}{{\partial}}\newcommand{\sig}{{\sigma}}
\newcommand{\Sig}{{\Sigma}}
\renewcommand{\th}{{\theta}}
\newcommand{\la}{{\lambda}}

\newcommand{\om}{{\omega}}\newcommand{\Om}{{\Omega}}

\makeatletter
\renewcommand\section{\@startsection{section}{1}{\z@}
  {-3.5ex \@plus -1ex \@minus -.2ex}%
    {2.3ex \@plus.2ex}
    {\vskip5pt\normalfont\large\bfseries}}
\renewcommand\subsection{\@startsection{subsection}{2}{\z@}%
    {-3.5ex \@plus -1ex \@minus -.2ex}%
    {1.3ex \@plus.2ex}%
    {\noindent\normalfont\normalsize\underline}}%
\makeatother

\numberwithin{equation}{section}

\begin{document}
\thispagestyle{empty}

\begin{center}
 
  {\huge Yang-Mills model for centrally extended 2d gravity}
  \vskip 25pt
     {\large
       Sara~Abentin\footnote{\href{mailto:sarabent@ucm.es}{sarabent@ucm.es},
         \href{https://orcid.org/0000-0002-1726-3793}{ORCID: 0000-0002-1726-3793}}
     and
     Fernando~Ruiz~Ruiz\footnote{\href{mailto:ferruiz@ucm.es}{ferruiz@ucm.es},
       \href{https://orcid.org/0000-0003-1571-2468}{ORCID: 0000-0003-1571-2468}}}
     \vskip 6pt
     \emph{Departamento de F\'{\i}sica Te\'orica, Facultad de Ciencias F\'{\i}sicas}\\
     \emph{Universidad Complutense de Madrid, 28040 Madrid Spain}

\end{center}
\vskip 30pt 

{\leftskip=30pt\rightskip=30pt 
  
  \noindent A Yang-Mills theory linear in the scalar curvature for 2d gravity
  with symmetry generated by the semidirect product formed with the Lie
  derivative of the algebra of diffeomorphisms with the two-dimensional
  Abelian algebra is formulated. As compared with dilaton models, the r\^ole
  of the dilaton is played by the dual field strength of a $U(1)$ gauge
  field. All vacuum solutions are found. They are either black holes or have
  constant scalar curvature. Those with constant scalar curvature have
  constant dual field strength. In particular, solutions with vanishing
  cosmological constant but nonzero scalar curvature exist.  In the
  conformal-Lorenz gauge, the model has a CFT interpretation whose residual
  symmetry combines holomorphic diffeomorphisms with a subclass of $U(1)$
  gauge transformations while preserving $\textnormal{dS}_2$ and
  $\textnormal{AdS}_2$ boundary conditions. This is the same symmetry as in
  Jackiw-Teitelboim-Maxwell gravity considered by Hartman and Strominger. It
  is argued that this is the only nontrivial Yang-Mills model linear in the
  scalar curvature that exists for real Lie algebras of dimension
  four.\\[5pt]%

  \noindent {\sc keywords:} 2d gravity, Abelian gauge field, dS/AdS boundary,
  CFT interpretation.

\par}

\vskip 35pt
\section{Introduction}

Two-dimensional dilaton gravity models provide effective theories to study
regimes of interest in higher-dimensional gravity. Among them,
are Jackiw-Teitelboim (JT) gravity~\cite{Teitelboim,Jackiw}, with a linear
coupling $\phi R$ between the dilaton and the scalar curvature and which
accounts for near-horizon theories in higher-dimensional near-extremal black
holes; the Almheiri-Polchinski~\cite{Almheiri-Polchinski} models, with
quadratic coupling $\phi^2 R$, that consistently explain the holographic flow
to $\textnormal{AdS}_2\times X$ of many theories; and the
Callan-Giddings-Harvey-Strominger model~\cite{CGHS}, with exponential coupling
$e^{-\phi} R$, that provides a 2d setting to analytically understand the
formation and subsequent evaporation of a black hole.

Here we propose a nondilaton model in which the r\^ole of the dilaton is
played by the dual field strength $\ast\/F$ of an Abelian gauge field
$A_\mu$. The model has classical action
\begin{equation}
  S=\frac{1}{2\kappa}\int d^2x\,\sqrt{|g|}\, \bigg( R \ast{\!F} 
  - \frac{1}{4\ell^2}\,F^2 + \frac{\gamma}{\ell^2}\bigg) \,,
  \label{action-intro}
\end{equation}
with $\ast\/F=\frac{1}{2}\epsilon^{\mu\nu}F_{\mu\nu}$ and
$F_{\mu\nu}=\partial_\mu\/A_\nu-\partial_\nu\/A_\mu$. The square $F^2$ stands
for $F^{\mu\nu}F_{\mu\nu}$, $\ell$ is a characteristic length, $\kappa$ and
$\gamma$ are dimensionless constants, and $A_\mu$ has dimensions of
length. The term $R\ast\!F$ couples the scalar curvature to a $U(1)$ gauge
field in an unusual fashion, with\, $\ast{F}$ a gravity source linear in the
gauge field. This point of view can be turned around to regard
$\epsilon^{\mu\nu}\pa_\nu\/R$ as a gauge current.

The idea that motivated this investigation was to formulate a 2d gravity model
as a Yang-Mills theory whose classical action is linear in the Ricci scalar.
In two dimensions, for a gauge symmetry generated by the 2d Poincar\'e
algebra, the resulting Yang-Mills action is quadratic in the scalar
curvature. However, as we discuss in Section~2, for the centrally extended
Poincar\'e algebra $\mathfrak{p}_1$, the Utiyama-Kibble-Sciama
approach~\cite{Utiyama,Kibble,Sciama}, modified along the lines of
refs.~\cite{Hartong-1,Hartong-2,Grumiller}, leads to the action $S$ above. The
modification consists in no longer considering plane gauge transformations but
a variant of them that can be understood as the semidirect product formed by
the Lie derivative of diffeomorphisms with Abelian gauge transformations. This
ensures that the zweibein postulate that maps the torsion and Riemann
curvature to the gauge field strengths is valid for arbitrary torsion.

Coming back to the dilaton picture, one may think of $S$ in the following
terms. Consider models with Lagrangian density\, ${\cal L}= \phi R+ V(\phi)$,
JT gravity corresponding to $V(\phi)=\gamma\phi/\ell^2$. The
action $S$ above is obtained by setting $\phi$ equal to $\ast{F}$ and taking
$V(\phi)=\gamma/\ell^2+\phi^2/2\ell^2$. This changes the field content, hence
the model itself, but leads to $S$. From this point of view, including in
$V(\phi)$ a linear term $\phi=\ast{F}$ contributes to the action with a total
derivative that we ignore.

The occurrence of the term $F^2$ in the action~(\ref{action-intro}) ensures
that the model has black hole solutions similar to those in 2d dilaton
gravities~\cite{LMK,GLMK,Witten-JT}. This is discussed in Section~3, in which
all vacuum solutions to the model are found. Besides black holes, we find
spacetimes with constant scalar curvature $R=R_0/\ell^2$ and constant dual
field strength $\ast\/F=F_0$, with $R_0$ and $F_0$ satisfying
$F_0^2+2F_0R_0-2\gamma=0$. For a given cosmological constant $\gamma$, both
$\textnormal{dS}_2$ and $\textnormal{AdS}_2$ are possible. Having one or the
other depends on the value of $F_0$. This scenario occurs even for zero
cosmological constant, $\gamma=0$, in which case $R_0=-F_0/2$. If the term
$F^2$ in the action $S$ is removed, the classical theory still makes sense but
then only vacuum solutions with constant scalar curvature exist, $R_0$ and
$F_0$ being related through $R_0F_0 =\gamma$.

We wish to study the model~(\ref{action-intro}) in relation with other 2d
gravity-Maxwell models in the literature.  A particularly interesting one has
been considered by Hartman and Strominger~\cite{HS}, who have added to the JT
Lagrangian a term $-F^2/4$.  This results in a JT-Maxwell model that has an
$\textnormal{AdS}_2$ vacuum solution for constant $\ast\/F=E$. After fixing
the conformal gauge for the metric, the model has a conformal field theory
(CFT) interpretation, with a residual symmetry that combines conformal
diffeomorphisms and gauge transformations and that is generated by a Witt
algebra. If matter is included so that the $\textnormal{AdS}_2$ background is
preserved at the boundary and if, upon quantization, the $U(1)$ matter current
becomes anomalous, the Witt algebra becomes a Virasoro algebra with nonzero
central charge. The model~(\ref{action-intro}) shares the same symmetry. Hence
we expect it to also allow for a central charge. This is shown in Section~4.

We close by arguing in Section~5 that the action $S$ is unique in the sense
that it is the only Yang-Mills action linear in the scalar curvature that can
be written for symmetries generated by semidirect products obtained from
four-dimensional real Lie algebras.

\section{Classical action and its symmetries}

\subsection{ Local symmetry}

The starting point in our analysis is the central extension
$\mathfrak{p}_{1}$ of the Poincar\'e algebra in two spacetime
dimensions, spanned by the generators $P_0$ and $P_1$ of translations, the
generator $J:=M_{01}$ of boosts, and a central element $Q$, with Lie bracket
\begin{equation}
  [P_0,P_1]=Q\,,\qquad [P_0,J]=P_1\,,\qquad [P_1,J]=P_0\,,
  \qquad [Q,P_a]=[Q,J]=0. 
  \label{11-algebra}
\end{equation}
Consider a Lie algebra valued 1-form
\begin{equation}
  B_\mu=e^a{\!}_\mu P_a  + \omega_\mu J + A_\mu Q\,,\qquad a=0,1\,,
  \label{A-form}
\end{equation}
whose components are the zweibein $e^a{\!}_\mu(x)$, the spin connection
$\om_\mu(x)$, and a 1-form $A_\mu(x)$. If we assign dimensions of
$(\textnormal{length})^{-1}$ to $P_0$ and $P_1$, then $J$ is dimensionless and
$Q$ has dimensions of $(\textnormal{length})^2$.  Taking $e^a{\!}_\mu$ to be
dimensionless, $\omega_\mu$ and $A_\mu$ carry respectively dimensions of
$(\textnormal{length})^{-1}$ and length. The corresponding 2-form field
strength
\begin{equation}
  G_{\mu\nu} = \pa_\mu\/B_\nu - \pa_\nu\/B_\mu + [B_\mu\,,B_\nu ]
  =: T^a{\!}_{\mu\nu}\, P_a + \Om_{\mu\nu}\, J + Z_{\mu\nu}\,Q\,,
  \label{F-form}
\end{equation}
has components 
\begin{gather}
  T^a{\!}_{\mu\nu} = \pa_\mu\/e^a{\!}_\nu - \pa_\nu\/e^a{\!}_\mu
    - \epsilon^a{}_b\,(\,\om_\mu\/e^b{\!}_\nu
    - \omega_\nu\/e^b{\!}_\mu \,)\,, \label{Ta}\\
    \Om_{\mu\nu}  = \pa_\mu\om_\nu -  \pa_\nu\om_\mu\,, 
           \label{Omega}\\
    Z_{\mu\nu}  = \pa_\mu\/A_\nu -  \pa_\nu\/A_\mu
     - \epsilon_{ab}\,e^a{\!}_\mu\,e^b{\!}_\nu\,,
      \label{Z}
\end{gather}
where we have used the conventions
\begin{equation}
     \epsilon^0{\!}_1=\epsilon^1{\!}_0=1,\qquad
    \epsilon_{ab}=\eta_{ac}\epsilon^{c}{\!}_b,\qquad
    \eta_{ab}=\textnormal{diag}(-1,+1).
\label{conventions}
\end{equation}

We next follow refs.~\cite{Hartong-1,Hartong-2} and, instead of conventional
gauge transformations, consider local transformations of the form
\begin{equation}
  {\dl}_{(\xi,\Sig)}B_\mu  = {\cal L}_\xi\/B_\mu
  +  \pa_\mu\Sig + [B_\mu,\Sig]\,.
  \label{delta-B}
\end{equation}
Here ${\cal L}_\xi$ is the Lie derivative along an arbitrary vector field
$\xi=\xi^\mu\pa_\mu$ generating the diffeomorphism
$x^\mu \to x^\mu+\xi^\mu(x)$, and
\begin{equation}
   \Sig=\th J + \tau Q
  \label{Sigma}
\end{equation}
is a function that takes values in the Abelian subalgebra spanned by $J$ and
$Q$, with $\th(x)$ and $\tau(x)$ arbitrary functions of dimensions 0 and
$(\textnormal{length})^2$.  Altogether there are four independent local
parameters, the two components of $\xi^\mu$ and the two functions $\th$ and
$\tau$. Under ${\dl}_{(\xi,\Sig)}$ the field strength $G_{\mu\nu}$ transforms
as
\begin{equation}
  {\dl}_{(\xi,\Sig)}G_{\mu\nu} = {\cal L}_\xi\/G_{\mu\nu}
  + [G_{\mu\nu},\Sig]\,.
  \label{delta-F}
\end{equation}
The transformation ${\dl}_{(\xi,\Sig)}$ is the combination
\begin{equation}
  {\dl}_{(\xi,\Sig)} = {\cal L}_\xi + \tilde{\dl}_\Sig
  \label{combination}
\end{equation}
of an arbitrary change of coordinates implemented by the Lie derivative
${\cal L}_\xi$, and a conventional gauge transformation generated by
$\tilde{\dl}_\Sig$.

The transformations ${\dl}_{(\xi,\Sig)}$ close an algebra, with closure
relation
\begin{equation}
  \big[\,{\cal L}_{\xi_1}\! +  \tilde{\dl}_{\Sig_1}\,,
       \,{\cal L}_{\xi_2} \!+ \tilde{\dl}_{\Sig_2}\,]
  = {\cal L}_{[\xi_1,\xi_2]} + 
    \tilde{\dl}_{{\cal L}_{\xi_2}\Sig_1-{\cal L}_{\xi_1}\Sig_2}\,.
  \label{closure}
\end{equation}
This Lie bracket can be described in mathematical terms as follows.  Consider
the Lie algebra ${\cal X}$ of vector fields on the spacetime manifold $M$.
and its representation provided by the Lie derivative, so that every vector
field $\xi$ is realized as a Lie derivative ${\cal L}_\xi$. The vector space
of all pairs $({\cal L}_\xi,\Sig) := {\cal L}_\xi\! + \tilde{\dl}_\Sig$
equipped with the bracket~(\ref{closure}) is a Lie algebra. It is, in fact, the
the semidirect product\, ${\cal X}\ltimes\mathfrak{a}_2$ \,of ${\cal X}$ with
the Abelian algebra $\mathfrak{a}_2={Span}\{J,Q\}$ formed with the Lie
derivative. The transformation laws of the zweibein, spin connection and
central gauge field are
\begin{gather}
  {\dl}_{(\xi,\Sig)}\,e^a_{\,\mu}= {\cal L}_\xi\/e^a_{\,\mu}
       + \epsilon^a_{\,\,b}\,e^b_{\,\mu}\, \th, \label{delta-e}\\[3pt]
  {\dl}_{(\xi,\Sig)}\,\om_\mu = {\cal L}_\xi\om_\mu + \pa_\mu\th,
       \label{delta-omega}\\[3pt]
  {\dl}_{(\xi,\Sig)}A_\mu = {\cal L}_\xi\/A_\mu + \pa_\mu\tau\,,
       \label{delta-A}
\end{gather}
whereas those of the field strength components take the form
\begin{gather}
  {\dl}_{(\xi,\Sig)}T^a{\!}_{\mu\nu} = {\cal L}_\xi\/T^a{\!}_{\mu\nu}
       + \epsilon^a{\!}_b\,T^b{\!}_{\mu\nu}\, \th,
      \label{delta-T}\\[3pt] 
  {\dl}_{(\xi,\Sig)}\Om_{\mu\nu} = {\cal L}_\xi\Om_{\mu\nu},
       \label{delta-Omega} \\[3pt]                                                     
  {\dl}_{(\xi,\Sig)}Z_{\mu\nu} =  {\cal L}_\xi\/Z_{\mu\nu}\,.
       \label{delta-Z}
\end{gather}

We next map the spin connection $\om_\mu$ to an affine connection
$\Gamma^\a{\!}_{\mu\nu}$ through the zweibein postulate
\begin{equation}
  {\cal D}_\mu\/e^a{\!}_\nu:=
  \pa_\mu\/e^a{\!}_\nu - \Gamma^\a{\!}_{\mu\nu}\,e^a{\!}_\a
  - \epsilon^a{\!}_b\,\om_\mu\,e^b{\!}_\nu = 0\,.
  \label{ZP}
\end{equation}
The derivative ${\cal D}_\mu\/e^a{\!}_\nu$ defined by the left-hand side of
this equation transforms under ${\dl}_{(\xi,\Sig)}$ as
\begin{equation}
  {\dl}_{(\xi,\Sig)}\big({\cal D}_\mu\/e^a{\!}_\nu\big)
  = {\cal L}_\xi\big({\cal D}_\mu\/e^a{\!}_\nu \big)
  + \epsilon^a{\!}_b\,\big({\cal D}_\mu\/e^b{\!}_\nu \big)\,\th\,, 
  \label{ZP-transformation}
\end{equation}
so that condition ${\cal D}_\mu\/e^a{\!}_\nu=0$ remains invariant. It is
precisely invariance under ${\dl}_{(\xi,\Sig)}$ that excludes terms in
eq.~(\ref{ZP}) of the form $c^a{\!}_b\,A_\mu\/e^a{\!}_\nu$ with nonzero
coefficients $c^a{\!}_b$. Using the solution to eq.~(\ref{ZP}) for
$\Gamma^\a{\!}_{\mu\nu}$ in terms of $\om_\mu$, the Riemann
$R^\a{\!}_{\b\mu\nu}$ and torsion $S^\a{\!}_{\mu\nu}$ tensors\footnote{We
  follow the convention\,
  $R^\a{\!}_{\b\mu\nu}= \pa_\mu\Gamma^\a{\!}_{\nu\b} +
  \Gamma^\a{\!}_{\mu\sig}\,\Gamma^\sig{\!}_{\nu\b}- (\mu\leftrightarrow\nu)$
  \,and\, $S^\a{\!}_{\mu\nu}=2\Gamma^\a{\!}_{[\mu\nu]}$.} are mapped to
$\Om_{\mu\nu}$ and $T^a{\,}_{\mu\nu}$ through
\begin{equation}
  R^\a{\!}_{\b\mu\nu} = 
   - E^\a{\!}_a\,\epsilon^a{\!}_b\,e^b{\!}_\b\, \Om_{\mu\nu}\,,\qquad
  S^\a{\!}_{\mu\nu} = E^\a{\!}_a\,T^a{\!}_{\mu\nu}\,.
  \label{Riemann}
\end{equation}
Here $E^\mu{\!}_a$ is the inverse zweibein, defined by
$E^\mu{\!}_ae^a{\!}_\nu=\dl^\mu{\!}_\nu$ and
$e^a{\!}_\mu\/E^\mu{\!}_b=\dl^a{\!}_b$.

\subsection{Comparison with conventional gauge transformations}

Under standard $\mathfrak{p}_{1}$ gauge transformations, the
1-form $B_\mu$ transforms as\,
$\tilde{\dl}_\Lambda\/B_\mu = \pa_\mu\Lambda + [B_\mu,\Lambda]$, with
$\Lambda=\rho^aP_a+\zeta\/J + \sigma\/Q$ an arbitrary gauge parameter
function. This gives for the components of $B_\mu$ the transformation laws
\begin{gather}
  \tilde{\dl}_\Lambda e^a{\!}_\mu = \pa_\mu\rho^a
  - \epsilon^a{\!}_b\,(\om_\mu\, \rho^b - e^b{\!}_\mu\,\zeta) ,
  \label{conventional-delta-e} \\[3pt]
  \tilde{\dl}_\Lambda \om_\mu =\pa_\mu\zeta ,
  \label{conventional-delta-omega} \\[3pt]
  \tilde{\dl}_\Lambda A_\mu = \pa_\mu\sigma - \epsilon_{ab}\,e^a{\!}_\mu\,
  \rho^b\,.
\label{conventional-delta-A}
\end{gather}
It is straightforward to check that there is not any zweibein postulate linear
in both $\om_\mu$ and $A_\mu$ that remains invariant under
$\tilde{\dl}_\Lambda$. Furthermore, standard arguments~\cite{Witten-3d} show
that eq.~(\ref{ZP}) remains $\tilde{\dl}_\Lambda$ invariant, modulo a
change of coordinates, only if the torsion vanishes. This suggests that, to
study scenarios with nonzero torsion, it is convenient to use the symmetry
${\dl}_{(\xi,\Sig)}$ rather than $\tilde{\dl}_{\Lambda}$. Transformations of
type ${\dl}_{(\xi,\Sig)}$ have been used in studies of
Horava-Lifshitz~\cite{Hartong-1} and Carrollian~\cite{Hartong-2}
gravities. The two transformations are related through~\cite{Grumiller}
\begin{equation}
  {\dl}_{(\xi,\Sig)} B_\mu = \tilde{\dl}_{\Lambda}\/ B_\mu + \xi^\nu
  G_{\nu\mu},  ~~~\textnormal{with}~~~   \Lambda=\xi^\mu\/B_\mu + \Sig\,.
\label{delta-delta}
\end{equation}
For $\dl_{(\xi,\Sig)}$ and $\tilde{\dl}_{\Lambda}$ to agree,
the torsion, and also $\Om_{\mu\nu}$ and $Z_{\mu\nu}$ must vanish.

\subsection{Invariant Lagrangian}

We are interested in Lagrangians that are invariant
under~${\dl}_{(\xi,\Sig)}$, linear in the Riemann curvature and at most
quadratic in first derivatives of the fields.  Because of eq.~(\ref{Riemann}),
linearity in the Riemann tensor is equivalent to linearity in $\Om_{\mu\nu}$.
In accordance with eqs.~(\ref{Ta})-(\ref{Z}), the most general Lagrangian of
this type is
\begin{equation}
  {\cal L} = \sqrt{|g|}\,\Big[
  c_1\ast\hspace{-1pt}\Om + \frac{c_2}{\ell^2}\ast{\!Z} + c_3\, \eta_{ab}\ast\/T^a \!\ast\/T^{b}
 + c_4 \ast{\Om}\ast{Z} +  \frac{c_5}{\ell^2}\,(\ast{Z})^2 \Big]\,,
  \label{L-0}
\end{equation}
where\, $c_1,\ldots,c_5$ \,are arbitrary constants and 
\begin{equation}
   \ast{\!\Phi}=\frac{1}{2}\,\eps^{\mu\nu}\Phi_{\mu\nu}
   \label{F-dual}
\end{equation}
is the dual of the 2-form $\Phi_{\mu\nu}$, with $\epsilon^{\mu\nu}$ the
antisymmetric pseudotensor.

In what follows we restrict ourselves to Levi-Civita connections, for which
the torsion vanishes,
\begin{equation}
  T^a{\!}_{\mu\nu}=0,
\label{zero-torsion}
\end{equation}
and the metric is given in terms of the zweibein by
\begin{equation}
  g_{\mu\nu}=\eta_{ab}\,e^a{\!}_\mu\,e^b{\!}_\nu\,.
  \label{LC}
\end{equation}
In this case, using eq.~(\ref{Riemann}) to write $\ast\/\Om$ in terms of the
Ricci scalar $R$, and introducing
\begin{equation}
  F_{\mu\nu}=\pa_\mu\/A_\nu -  \pa_\nu\/A_\mu\,,
  \label{Fmunu}
\end{equation}
we have 
\begin{equation}
  \ast\Om = -\frac{1}{2}\,R=\nabla_{\!\mu}(\epsilon^{\mu\nu}\om_\nu)\,,
  \qquad   \ast\/Z = \ast\/F + 1 =\nabla_{\!\mu}(\epsilon^{\mu\nu}A_\nu) +1 \,.
 \label{duals-Omega-Z}
\end{equation}
Substituting these equations in ${\cal L}$ above and discarding total
derivatives, we are left with
\begin{equation}
  {\cal L} = \sqrt{|g|}\,\Big[\,
  \frac{c_2+c_5}{\ell^2} - \frac{c_4}{2}\,R\ast{\!F}
  +  \frac{c_5}{\ell^2} \,(\ast{F})^2\, \Big]\,.
  \label{L-1}
\end{equation}
Making the change $A_\mu\to -(c_4/4c_5) A_\mu$, and setting
$\kappa=4c_5 /c_4^2$ and $\gamma/2\kappa=c_2+c_5$, we arrive at the
classical action
\begin{equation}
  S=\frac{1}{2\kappa}\int d^2x\,\sqrt{|g|}\, \bigg[ R \ast{\!F} 
  + \frac{1}{2\ell^2}\,(\ast F)^2 + \frac{\gamma}{\ell^2}\bigg] 
  + S_m .
  \label{action}
\end{equation}
This is the action in eq.~(\ref{action-intro}), for in two dimensions with
Lorentzian signature one has
\begin{equation}
  \epsilon^{\mu\nu}\epsilon^{\a\b}=g^{\mu\b}g^{\nu\a}- g^{\mu\a}g^{\nu\b}\,.
  \label{identity-epsilon}
\end{equation}
In eq.~(\ref{action}) we have included a matter contribution $S_m$ that couples
$g_{\mu\nu}$ and $A_\mu$ to other fields but does not contain derivatives of
$g_{\mu\nu}$ and $A_\mu$.

\section{Vacuum solutions}

Varying the action with respect to $g_{\mu\nu}$ and using that in two
dimensions\, $2\,\dl_g(\ast\/F)=\ast\/F\,\dl\/g^{\mu\nu}g_{\mu\nu}$ \,and\,
$2R_{\mu\nu}=g_{\mu\nu}R$, one has
\begin{align}
  \dl_gS &
            = \frac{1}{2\kappa}\int\! d^{2\!}x\,\sqrt{|g|}\,
    \nabla_{\!\mu} \Big[ \nabla_{\!\nu}\big(v^{\mu\nu\!}\ast{\!F}\,\big)
           - 2\,v^{\mu\nu}\,\nabla_{\!\nu\!}\left(\ast{F}\right)\Big]
  \nonumber\\[4.5pt]
  & + \frac{1}{2}\int\! d^{2\!}x\,\sqrt{|g|}\,
    \dl\/g^{\mu\nu}\,\left( T^{g}_{\mu\nu} + T^{m}_{\mu\nu} \right) \,.
    \label{sin-0}
\end{align}
The first term is a boundary term, with $v^{\mu\nu}$ given by
\begin{equation}
  v^{\mu\nu} =-\dl\/g^{\mu\nu} + g^{\mu\nu}\,\dl\/g^{\a\b}g_{\a\b},
  \label{sin-1}
\end{equation}  
$T^{g}_{\mu\nu}$ has the form
\begin{equation}
  \kappa T^{g}_{\mu\nu} = \frac{1}{2}\,g_{\mu\nu}\bigg[
  R\ast{\!F} +\frac{(\ast{F})^2}{2\ell^2} 
  +2\,\nabla^2\!\ast{\!F}  - \frac{\gamma}{\ell^2} \bigg]
  -\nabla_{\!\mu}\nabla_{\!\nu\!}\ast{\!F}  ,
  \label{sin-2}
\end{equation}
and $ T^m_{\mu\nu}$ is the matter energy-momentum tensor,
\begin{equation}
  T^m_{\mu\nu} =  \frac{2}{\sqrt{|g|}}\,
  \frac{\dl\/S_m}{\dl\/g^{\mu\nu}} \, .
\end{equation}
Variation of $S$ with respect to $A_\mu$ yields in turn
\begin{align}
  \dl_A\/S & = \frac{1}{\kappa}\int\! d^{2\!}x\,\sqrt{|g|}\;
   \nabla_{\!\mu}\Big(\, \epsilon^{\mu\nu}\,\epsilon^{\a\b}
   \left[\, \nabla_{\!\a}\big( C_\b\,\dl{\hspace{-0.9pt}}A_\nu\big)
      - C_\b \,\nabla_{\!\a}\dl{\hspace{-0.9pt}}A_\nu \right]\Big)
      \nonumber \\[4.5pt]
   & + \int\! d^{2\!}x\,\sqrt{|g|}\;
     \dl{\hspace{-0.9pt}}A_\mu\, \left[  \frac{1}{2\kappa}\,
     \epsilon^{\mu\nu}\,\pa_\nu
    \Big( R+ \frac{\ast{F}}{\ell^2}\Big)  + J^{m\mu} \right] \,,\label{sin-3}
\end{align}
where $C_\mu$ reads
\begin{equation}
C_\mu = -\,\om_\mu +\frac{A_\mu}{2\ell^2}\,
\label{C-mu}
\end{equation}
and $J^{m\mu}$ is the $U(1)$ matter current
\begin{equation}
 J^{m\mu} = \frac{1}{\sqrt{|g|}}\,\frac{\dl S_m}{\dl A_\mu}\,.
\label{J-mu-matter}
\end{equation}

We assume suitable boundary conditions, so that the boundary terms in
eqs.~(\ref{sin-0}) and~(\ref{sin-3}) vanish. The field equations are then
\begin{equation}
T^g_{\mu\nu}+T^m_{\mu\nu}=0
  \label{T-mu-nu}
\end{equation}
and 
\begin{equation}
  J_{\scriptscriptstyle T}^\mu:=
  \frac{1}{2\kappa}\,\epsilon^{\mu\nu}\nabla_\nu
  \Big(R+\frac{\ast F}{\ell^2}\Big) +J^{m\mu}=0\,.
  \label{J-total-mu}
\end{equation}
Acting on eq.~(\ref{J-total-mu})  with $\nabla_{\mu}$ and using
$\nabla_{\mu}\epsilon^{\mu\nu}=0$, we have
\begin{equation}
  \nabla_{\mu}J^{m\mu}=0\,.
  \label{conservation-J}
\end{equation}
Hence the matter contribution to the $U(1)$ gauge current must be conserved. 

We wish to solve the field equations in vacuum. This is most conveniently done
in the conformal gauge with light-cone coordinates
\begin{equation}
  ds^2= -\,e^{2\varphi}\,dx^+\,dx^-,\qquad   x^\pm=t\pm x\,,
  \label{conformal-gauge}
\end{equation}
in which the equations take the form
\begin{align}
  \pa_\pm\pa_\pm\!\ast\!F - 2\,\pa_\pm\varphi\;\pa_\pm\!\ast\!F
  & =0 , \label{vac-field-pmpm} \\[4.5pt]
  2\,\pa_+\pa_-\!\ast\!F - \frac{1}{2}\,e^{2\varphi}\Big[ R\ast\!F
  +\frac{(\ast F)^2}{2\ell^2} -\frac{\gamma}{\ell^2} \Big]
  &= 0, \label{vac-field-eqn-pm} \\[3pt]
  \pa_\pm\Big(R+\frac{1}{\ell^2}\ast\!F\Big)
  & =0 , \label{vac-field-eqn-A}
\end{align}
and the Ricci scalar is given by
\begin{equation}
  R= 8\,e^{-2\varphi}\,\pa_+\pa_-\varphi\,.
\label{Liouville}
\end{equation}
Equation~(\ref{vac-field-eqn-A}) can be regarded as an integrability condition, for it
is reproduced by acting with $\pa_\mp$ on eq.~(\ref{vac-field-pmpm}) and using
eq.~(\ref{vac-field-eqn-pm}). It ensures that the boundary term in
eq.~(\ref{sin-3}) vanishes, since the latter can also be
written as
\begin{equation}
\frac{1}{\kappa}\int\! d^{2\!}x\,\sqrt{|g|}\;
   \nabla_{\!\mu}\Big[\, \epsilon^{\mu\nu}
   \left(R+\frac{\ast\/F}{\ell^2} \right) \dl\/A_\nu\Big]\,.
   \label{sin-4}
 \end{equation}
 To solve eqs.~(\ref{vac-field-pmpm})-(\ref{vac-field-eqn-A}), we
 distinguish between constant and nonconstant scalar curvature.

\subsection{Solutions with constant scalar curvature.}

If $R$ is constant, eq.~(\ref{vac-field-eqn-A}) implies that so is $\ast
F$. We thus write
\begin{align}
  \bar{R}&= \frac{R_0}{\ell^2}\,,\label{sol-R}\\[3pt] 
    \ast{\bar{F}}&=F_0 \label{sol-F}\,,
\end{align}
with $R_0$ and $F_0$ dimensionless constants satisfying the constraint
provided by eq.~(\ref{vac-field-eqn-pm}),
\begin{equation}
   F_0^2+2F_0R_0-2\gamma=0\,.
  \label{restriction}
\end{equation}
Vacuum spacetime is locally isomorphic to Minkowski, $\textnormal{dS}_2$, or
$\textnormal{AdS}_2$. Equation~(\ref{sol-F}) can be recast as
\begin{equation}
  \bar{F}_{+-} = -\frac{F_0}{2}\,\,e^{2\bar{\varphi}}\,,
\label{sol-F+-}
\end{equation}
with $\bar{\varphi}$ a solution to the Liouville equation (\ref{Liouville})
for $R=R_0/\ell^2$. An expression for the gauge potential solution
$(\bar{A}_+,\bar{A}_-)$ can be found by choosing a gauge and solving
eq.~(\ref{sol-F}). Here we will work in the Lorenz gauge
\begin{equation}
  \pa_+\/A_-+ \pa_-\/A_+ =0\,,
  \label{Lorenz-condition}
\end{equation}
in which $A_\pm$ and $F_{+-}$ become~\cite{HS}
\begin{align}
  A_\pm & =  \mp \pa_\pm\/a \,, \label{A-Lorenz}\\[3pt]
  F_{+-} & = 2\pa_+\pa_-\/a\,,\label{F-Lorenz}
\end{align}
with $a=a(x^+,x^-)$ an arbitrary function of its arguments with dimensions of
$(\textnormal{length})^2$. Upon substitution in eq.~(\ref{sol-F}), we obtain
\begin{equation}
     4e^{-2\bar{\varphi}}\,\pa_+\pa_-\bar{a}= - F_0
  \label{eq-a-Lorenz}
\end{equation}
We note that $a$ plays a r\^ole similar to that of $\varphi$. In fact, solving
the vanishing torsion equations $T^a_{\,\,+-}=0$ for the spin connection, we
have
\begin{equation}
  \om_\pm= \mp \pa_\pm\varphi \,, 
  \label{conformal-spin-connection}
\end{equation}
which has the same form as eq.~(\ref{A-Lorenz}).

For zero scalar curvature, $R_0=0$, the vacuum spacetime is locally
isomorphic to Minkowski space, metric
$ds^2_{\scriptscriptstyle R_0=0} =-dx^-dx^+$. In this case,
$\bar{\varphi}=0$, and eq.~(\ref{eq-a-Lorenz}) becomes
$\pa_+\pa_-\bar{a}= -F_0/4$, with $F_0=\pm\sqrt{2\gamma}$, which requires
$\gamma>0$. The solution for ${a}$ is then
\begin{equation}
  \bar{a}_{\scriptscriptstyle R_0=0} = 
  - \frac{F_0}{8}\,(x^++x^-)^2 + f_{\scriptscriptstyle R}(x^+)
  + f_{\scriptscriptstyle L}(x^-)\,,
   \label{flat-sol-A}
\end{equation}
where $f_{\scriptscriptstyle R}(x^+)$ and $f_{\scriptscriptstyle L}(x^-)$ are
arbitrary functions of their arguments with dimensions of
$\textnormal{length}^2$. The arbitrariness in
$f_{\scriptscriptstyle R}$ and $f_{\scriptscriptstyle L}$ is reminiscent of
the fact that the Lorenz condition~(\ref{Lorenz-condition}) does not
completely eliminate gauge invariance but leaves a residual gauge symmetry.

If $R_0\neq 0$, the general solution to eq.~(\ref{eq-a-Lorenz}) is given in
terms of the solution $\bar{\varphi}$ to Liouville's
equation~(\ref{Liouville}) by
\begin{equation}
 \bar{a}_{\scriptscriptstyle R_0\neq 0}=-\frac{2F_0\ell^2}{R_0}\,\bar{\varphi}
      + f_{\scriptscriptstyle R}(x^+) + f_{\scriptscriptstyle L}(x^-)\,. 
  \label{sol-a-Lorenz}
\end{equation}
where $F_0/R_0$ on the right-hand side is the solution to
eq.~(\ref{restriction}),
\begin{equation}
  \frac{F_0}{R_0} = -1\pm\sqrt{1+\frac{2\gamma}{R_0^2}}\,.
  \label{restriction-solution}
\end{equation}
These solutions are different from those of JT gravity. In our case, the
Ricci scalar $R_0/\ell^2$ is no longer equal to $-\gamma/\ell^2$. For a given
value of $\gamma$ such that $R_0^2+2\gamma>0$, the scalar curvature may be
positive or negative, depending on $F_0$. We remark that if the term
$(\ast{F})^2$ is removed from the classical action, the vacuum solutions are
the same, the only difference being that now $F_0R_0=\gamma$. Furthermore, for
vanishing cosmological constant, $\gamma=0$, and provided the term
$(\ast{F})^2$ is kept, vacuum spacetime will be nonflat with constant scalar
curvature $R=-F_0/2\ell^2$. In particular, a gauge field with $F_0=\mp 4$ will
generate a $\textnormal{dS}_2/\textnormal{AdS}_2$ with scalar curvature
$R_0=\pm 2$.

Coming back to the case of arbitrary $\gamma$, for $R_0>0$, vacuum spacetime
is locally isomorphic to $\textnormal{dS}_2$, whose metric in Poincar\'e
coordinates\, $\{t>0,x\}$ \,is
\begin{equation}
  ds^2_{\scriptscriptstyle{dS}}= \frac{\ell^2}{t^2}\,(-dt^2+dx^2) 
   = -\,\frac{4\ell^2}{(x^++x^-)^2}\,dx^+dx^-\,.
\label{dS-metric}
\end{equation}
In these coordinates, $R_0=2$, and $\varphi$ becomes
\begin{equation}
  \bar{\varphi}_{\scriptscriptstyle dS} = \ln\Big(\frac{2\ell}{x^++x^-}\Big)\,.
  \label{dS-varphi}
\end{equation}
The expression of $\bar{a}_{\scriptscriptstyle dS}$, is obtained upon
substitution in eq.~(\ref{sol-a-Lorenz}). To eliminate the arbitrariness in
$f_{\scriptscriptstyle R}$ and $f_{\scriptscriptstyle L}$, we impose that the
component $C_t$ of $C_\mu$ in eq.~(\ref{C-mu}) vanishes at the boundary $t=0$,
\begin{equation}
  0= C_t\big\vert_{t=0} = -(\om_+ +\om_-)
      + \frac{1}{2\ell^2}\,(A_+ +A_-)\bigg\vert_{t=0}\,.
  \label{dS-C}
\end{equation}
This fixes $f_{\scriptscriptstyle R}$ and $f_{\scriptscriptstyle L}$ and gives
\begin{equation}
  \bar{a}_{\scriptscriptstyle dS}
  = \frac{2F_0\ell^2}{R_0}\,\ln\Big(\frac{x^++x^-}{2\ell}\Big)
 + \ell\a_1(x^++x^-) + \ell^2\a_0\, ,
  \label{dS-a-fixed}
\end{equation}
with $\a_0$ and $\a_1$ arbitrary dimensionless constants. The spin connection
and the gauge field are found upon substitution in
eqs.~(\ref{conformal-spin-connection})
and~(\ref{A-Lorenz}). Condition~(\ref{dS-C}) and the fact that
$\ast\bar{F}_{\scriptscriptstyle dS}$ is constant ensure that the boundary
terms in $\dl_gS$ and $\dl_AS$ vanish on shell.

For $R_0<0$, vacuum spacetime is locally isomorphic to $\textnormal{AdS}_2$,
with metric
\begin{equation}
  ds^2_{\scriptscriptstyle AdS}= \frac{\ell^2}{x^2}\,(-dt^2+dx^2)
   = -\,\frac{4\ell^2}{(x^+-x^-)^2}\,dx^+dx^-
\label{AdS-metric}
\end{equation}
in Poincar\'e coordinates\, $\{t,x>0\}$. Now $R_0=-2$, and
\begin{equation}
  \bar{\varphi}_{\scriptscriptstyle AdS}=\ln\!\Big(\frac{2\ell}{x^+-x^-}\Big),
  \qquad 
  \bar{a}_{\scriptscriptstyle AdS} = -\,\frac{2F_0\ell^2}{R_0}\,
  \bar{\varphi}_{\scriptscriptstyle AdS} + \ell\a_1(x^+-x^-)  + \ell^2\a_0
\label{AdS-varphi-a}
\end{equation}
for a boundary condition
\begin{equation}
  0= C_x\big\vert_{x=0} = -(\om_+ -\om_-)
      + \frac{1}{2\ell^2}\,(A_+ - A_-)\bigg\vert_{x=0}\,.
  \label{AdS-C}
\end{equation}
  
\subsection{Solutions with nonconstant scalar curvature: black holes}

To find the vacuum solutions with nonconstant scalar curvature, we employ
similar methods to those used in the proof of Birkhoff's theorem in 2d dilaton
gravity in refs.~\cite{LMK} and~\cite{GLMK}. Combine
eqs.~(\ref{vac-field-pmpm}) and ~(\ref{vac-field-eqn-A}) to write\,
\hbox{$\pa_\pm\big(e^{-2\varphi}\,\pa_\pm\/R\big)=0$}. This implies that
\begin{equation}
  \pa_+R =e^{2\varphi}\,h_{\scriptscriptstyle L}(x^-)\,,\qquad
  \pa_- R=e^{2\varphi}\,h_{\scriptscriptstyle R}(x^+)\,,
                    \label{proof-1-a}
\end{equation}
with $h_{\scriptscriptstyle L}(x^-)$ and $h_{\scriptscriptstyle R}(x^+)$
arbitrary functions of their arguments.  After having fixed the conformal
gauge, the model is still invariant under diffeomorphisms\,
\hbox{$x^+\!\to\tilde{x}^+(x^+)$} \,and\, $x^-\!\to\tilde{x}^-(x^-)$, under
which $h_{\scriptscriptstyle L}$ and $h_{\scriptscriptstyle R}$ transform as
\begin{equation}
  \tilde{h}_{\scriptscriptstyle L}(\tilde{x}^-)
  = h_{\scriptscriptstyle L}(x^-)\; \frac{d\tilde{x}^-}{dx^-}\,,\qquad
  \tilde{h}_{\scriptscriptstyle R}(\tilde{x}^+)
  = h_{\scriptscriptstyle R}(x^+)\; \frac{d\tilde{x}^+}{dx^+}\,.
\label{proof-4}
\end{equation}  
Use this residual symmetry to choose coordinates $\{\tilde{x}^+,\tilde{x}^-\}$
defined as the solutions to the equations
\begin{equation}
  \frac{d\tilde{x}^+}{dx^+}
    = \frac{1}{|h_{\scriptscriptstyle R}(x^+)|} \;, \qquad
  \frac{d\tilde{x}^-}{dx^-} = \frac{1}{|h_{\scriptscriptstyle L}(x^-)|}\;.
    \label{proof-3}
\end{equation}
For nonconstant curvature, $h_{\scriptscriptstyle L\!}$ and
$h_{\scriptscriptstyle R\!}$ are different from zero, so this change is
locally well defined. In the new coordinates, eqs.~(\ref{proof-1-a}) become
\begin{equation}
  \textnormal{sign}(h_{\scriptscriptstyle L})\;\widetilde{\!\pa_+ R}
  = \textnormal{sign}(h_{\scriptscriptstyle R})\; \widetilde{\!\pa_- R}
  = e^{2\tilde{\varphi}},
    \label{proof-4}
\end{equation}
It follows that either (i) $\tilde{\varphi}(\tilde{x})$ is a function of\,
$\tilde{x}=(\tilde{x}^+\!-\tilde{x}^-)/2$ or (ii) it is a function
$\tilde{\varphi}(\tilde{t})$ of\, $\tilde{t}=(\tilde{x}^+\!+\tilde{x}^-)/2$.

Let us consider scenario (i). In this case, $\varphi,\,R$ and $\ast F$ are
also functions of~$x$, where, to ease the writing, we have removed the tildes
from the notation.  Upon making the change $x \to r(x)$, with\,
$dr=e^{2\varphi(x)}dx $, the metric takes the form
\begin{equation}
  ds^2= -\,f(r)\,dt^2  +  \frac{dr^2}{f(r)} \,.
  \label{proof-5}
\end{equation}  
The function $f(r)$ is given in terms of $\varphi$ by
$f(r)=e^{2\varphi(x(r))}$. The scalar curvature becomes $R=-f''(r)$ and the
field equations~(\ref{T-mu-nu})-(\ref{J-total-mu}) read
\begin{align}
 (\ast\/ F)''=0\,, \label{proof-6-a} \\[4.5pt]
  f' (\ast F)' - f''\ast\!F
  + \frac{(\ast\/F)^2}{2\ell^2} -\frac{\gamma}{\ell^2} = 0\,,
  \label{proof-6-b} \\[4.5pt]
  -f''' + \frac{(\ast\/F)'}{\ell^2}=0\,,
  \label{proof-6-c}
\end{align}
where the prime denotes differentiation with respect to $r$. The solution to
eq.~(\ref{proof-6-a}) is $\ast F=a_1(r/\ell) + a_0$, with $a_1$ and $a_0$
integration constants. We are interested in $a_1\neq 0$, since $a_1\!=0$
corresponds to constant scalar curvature. As in 2d dilaton
gravity~\cite{Witten-JT}, we use the invariance of the metric under
$(t,r,f)\to(t/b_1,\,b_1r-b_0\ell,fb_1^2)$ to set $a_0=0$ and $a_1=1$. This
gives
\begin{equation}
  \ast F = \frac{r}{\ell}\,.
  \label{bh1}
\end{equation}
Equations~(\ref{proof-6-b}) and~(\ref{proof-6-c}) then yield
\begin{equation}
  f(r) = \frac{r^3}{6\ell^3}
  + c_0\,\frac{r^2}{2\ell^2}
  +\gamma\,\frac{r}{\ell} + c_1\,,
  \label{bh2}
\end{equation}
with $c_0$ and $c_1$ dimensionless integration constants. Being a cubic
polynomial with real coefficients, $f(r)$ has at least one real root. Call
$r_H$ to its largest real root. Since $f(r)$ is positive for $r>r_H$ and
changes its sign at $r=r_H$, the solution~(\ref{proof-5}), with $f$ in
eq.~(\ref{bh2}), can be understood as a black hole with horizon at $r_H$. Note
that $\pa_t$ is a timelike Killing vector for $f(r)>0$. Note also that 
the term $(\ast F)^2$ in the classical action is necessary to have solutions of
this type; otherwise the contribution $(\ast F)^2$ in
eq.~(\ref{proof-6-b}) is absent, and eq.~(\ref{proof-6-c}) reduces to $f'''=0$,
equivalently constant scalar curvature.

The other solution to eqs.~(\ref{proof-4}),
$\varphi(t)$ only depends on $t$, is analyzed similarly. After
reparametrizing $t$, and setting $r=x$, the dual field strength is now
$\ast F=-t/\ell$, and the metric takes the form
\begin{equation}
  ds^2= -\,\frac{dt^2}{f(t)}  +  f(t)\,dx^2 \,,
  \label{bh3}
\end{equation}
with
\begin{equation}
 f(t) = \frac{t^3}{6\ell^3}
  + d_0\,\frac{t^2}{2\ell^2}
  +\gamma\,\frac{t}{\ell} + d_1\,,
  \label{bh4}
\end{equation}
and $d_0$ and $d_1$ dimensionless constants of integration. For
$d_0=c_0$,
$d_1=c_1$, this metric describes the interior of the black
hole~(\ref{proof-5}), since when going across the horizon
$r_H$ of ~(\ref{proof-5}), the coordinate
$r$ becomes timelike and the metric can be cast as in eqs.~(\ref{bh3})
and~(\ref{bh4}).

\section{Boundary CFT description of the model}

In this section we present a CFT interpretation of the vacuum solutions with
constant scalar curvature. The classical action~(\ref{action}) in the
conformal-Lorenz gauge takes the form
\begin{equation}
  S_{\scriptscriptstyle CFT}= \frac{1}{\kappa} 
  \int\! dx^+dx^-\, \bigg[ 
    -8\,e^{-2\varphi}\,(\pa_+\pa_-a)\,\pa_+\pa_-\varphi 
     + \frac{2}{\ell^2} \,e^{-2\varphi}\,{(\pa_+\pa_-\/a)}^2
    + \frac{\gamma}{4\ell^2}\,e^{2\varphi}\bigg] + S_{m}\,.
\label{action-conformal}
\end{equation}
This action contains second derivatives with respect to time of $\varphi$
and $a$. It is invariant under conformal diffeomorphisms
\begin{equation*}
  x^\pm\to x^\pm+\xi^\pm(x^\pm)
  \label{conformal}
\end{equation*}
generated by arbitrary vector fields $\xi^+(x^+)\pa_+$ and $\xi^-(x^-)\pa_-$,
provided $e^\varphi$ transforms as a conformal field of weights
$({\scriptstyle 1/2,~1/2})$ and $a$ as a scalar. $S$ is also invariant under
residual gauge transformations
$a\to a+\tau_{\scriptscriptstyle R}(x^+)+\tau_{\scriptscriptstyle L}(x^-)$,
with $\tau_{\scriptscriptstyle R}(x^+)$ and $\tau_{\scriptscriptstyle L}(x^-)$
arbitrary functions of their arguments with dimensions of
$(\textnormal{length})^2$. Let us see that the combination of these two
symmetries is a residual symmetry $\dl_r$ of ${\dl}_{(\xi,\Sig)}$ specified by
$\xi^+$ and $\xi^-$.

\subsection{Residual symmetry}

In the conformal gauge, the zweibein is given by
\begin{equation}
  e^0{\!}_\pm = \frac{e^\varphi}{2}\,,\qquad 
  e^1{\!}_\pm  = \pm \frac{e^\varphi}{2}\,.
  \label{conformal-zweibein}
\end{equation}
To find ${\dl}_r\varphi=\dl_{(\xi,\Sig)}\varphi$ for a local parameter
\begin{equation}
  \xi_r=\xi^+\pa_+ + \xi^-\pa_-\,,
  \label{xi-res}
\end{equation}
we substitute the expressions~(\ref{conformal-zweibein}) in eq~(\ref{delta-e})
and use that ${\dl}e^\varphi=e^\varphi{\dl}\varphi$. This yields a system of
two equations for $\dl_r\varphi$ and $\th_r$, whose only solution is
\begin{gather}
  \dl_{r}\varphi = \big(\xi^+\pa_+
       + \xi^-\pa_-\big)\,\varphi 
     +\frac{1}{2}\,\big( \pa_+\xi^++\pa_-\xi^-\big)\,,
  \label{delta-varphi-res} \\[3pt]
  \th_r=\frac{1}{2}\,
      \big(\pa_-\xi_r^-\! - \pa_+\xi_r^+\big)\,.
  \label{theta-res}
\end{gather}
In eq.~(\ref{delta-varphi-res}) one recognizes the variation under conformal
diffeomorphisms of a field $e^\varphi$ with conformal weights
$({\scriptstyle 1/2,\,1/2})$.  Substituting the result~(\ref{theta-res}) for
$\th_r$ in the variation~(\ref{delta-omega}) of the spin connection, we have
\begin{equation}
  {\dl}_{r}\om_\pm
  = {\cal L}_{\xi_r}\om_\pm \mp\frac{1}{2}\, \pa_\pm^2\xi^\pm \,.
   \label{delta-omega-res}
 \end{equation}
The Lie derivative 
\begin{equation}
  {\cal L}_{\xi_r}\om_\pm =\left( \xi^+ \pa_+ + \xi^- \pa_-\right)\om_\pm  
  + \left(\pa_\pm\xi^\pm\right)\om_\pm 
  \label{Lie-omega}
\end{equation}
on the right-hand side accounts for the variation under conformal
diffeomorphisms of the 1-form $(\om_+,\om_-)$, while\,
$\mp\frac{1}{2}\, \pa_\pm^2\xi^\pm$ adds a $U(1)$ contribution generated by
boosts. The transformation law~(\ref{delta-omega-res}) can also be obtained by
using eq.~(\ref{delta-varphi-res}) in the solution $\om_\pm=\mp\pa_\pm\varphi$
to the vanishing torsion condition.

To find ${\dl}_{r}a$, set $\xi=\xi_r$ and $A_\pm=\mp\pa_\pm\/a$ in the
variations ${\dl}A_\pm$ in eq.~(\ref{delta-A}). This provides two equations
for ${\dl}_r\/a$ and $\tau_r$, whose solutions are
\begin{gather}
  {\dl}_ra = (\xi^+\pa_+ + \xi^-\pa_-)a + \tau_{\scriptscriptstyle R}(x^+)
  + \tau_{\scriptscriptstyle L}(x^-)\,, \label{bar-delta-a-off}\\[3pt]
 \tau_r=  \tau_{\scriptscriptstyle L}(x^-)
     - \tau_{\scriptscriptstyle R}(x^+)\,, \label{tau-res-off}
\end{gather}
with $\tau_{\scriptscriptstyle R}(x^+)$ and $\tau_{\scriptscriptstyle L}(x^-)$
arbitrary functions of their arguments.

To determine $\tau_{\scriptscriptstyle R}$ and $ \tau_{\scriptscriptstyle L}$,
one may proceed as follows. Regard any of the vacuum solutions
$\textnormal{dS}_2$ or $\textnormal{AdS}_2$ of Section~2 as the boundary of a
model with matter. Demanding the residual symmetry to be consistent with the
boundary, and recalling that at the boundary ${a}$ and ${\varphi}$ are related
through eq.~(\ref{sol-a-Lorenz}), it is straightforward that
\begin{equation}
  \tau_{\scriptscriptstyle R}(x^+) = - \frac{F_0\/\ell^2}{R_0}\, 
  \pa_+\xi^+,\qquad
  \tau_{\scriptscriptstyle L}(x^-) = - \frac{F_0\/\ell^2}{R_0}\,
  \pa_-\xi^-,
 \label{tau-res}
\end{equation}
and 
\begin{equation}
  \dl_r{a} = (\xi^+\pa_+ + \xi^-\pa_-)\,{a} 
    - \frac{F_0\/\ell^2}{R_0}\,\big(
    \pa_+\xi^+ + \pa_-\xi^- \big) \,.
    \label{delta-a-res}
\end{equation}
The variations $\dl_rA_\pm$ then read
\begin{equation}
  \dl_r{A_\pm} = {\cal L}_{\xi_r} A_\pm
     \pm\frac{F_0\ell^2}{R_0}\,\pa_\pm^2\,\xi^\pm\,.
    \label{delta-A-res}
\end{equation}
Residual transformations $\dl_r$ are thus determined by the vector field
$\xi_r=(\xi^+,\xi^-)$. We remark that $(R_0/2F_0\ell^2)a$ and
$(R_0/2F_0\ell^2)A_\pm$ transform under $\dl_r$ as $\varphi$ and $\om_\pm$.

\subsection{Witt algebra}

Denote by $\dl_r^+$ and $\dl_r^-$ the generators of the residual symmetries
associated to $\xi^+\pa_+$ and $\xi^-\pa_-$. Assume that $\xi^+(x^+)$ and
$\xi^-(x^-)$ can be expanded in power series of $x^+$ and $x^-$ with
coefficients $c_{n,+}$ and $c_{n,-}$, so that
\begin{equation}
  \xi^\pm\pa_\pm =\sum_n c_{n,\pm}\,(x^\pm)^{n+1}\pa_\pm\,.
  \label{analytic}
\end{equation}
In accordance with eqs.~(\ref{theta-res}) and~(\ref{tau-res}), the variation
${\dl}_r$ can be written as
\begin{equation}
  {\dl}_r =  {\dl}_r^+ +  {\dl}_r^-
  = \sum_n \big( c_{n,+}\,{\dl}^+_{n} + c_{n,-}\,{\dl}^-_{n}\big) \,,
\label{sum-dl}
\end{equation}
with ${\dl}^\pm_{n}$ the ${\dl}_{(\xi,\Sig)}$ transformation with parameters
\begin{equation}
  \xi_{n,\pm} = (x^\pm)^{n+1}\pa_\pm \,,
  \qquad \th_{n,\pm} = \mp\,\frac{n+1}{2}\,(x^\pm)^n\,, \qquad
   \tau_{n,\pm} = \pm  \frac{F_0\/\ell^2}{R_0}\,(n+1)\,(x^\pm)^n\,.
  \label{dl-n}
\end{equation}
The closure relation~(\ref{closure}) then implies
\begin{equation}
  \big [{\dl}^\pm_{n}\,,\,{\dl}^\pm_{m}\big] = (n-m)\;{\dl}^\pm_{n+m}\,.
  \label{dl-commutator-nm}
\end{equation}
The residual symmetry is hence generated by a Witt algebra.

The variation $\dl_r$ coincides with the combination of conformal and gauge
transformations introduced in JT-Maxwell gravity~\cite{HS}.

\subsection{Check of invariance of $\textnormal{dS}_2$ and $\textnormal{AdS}_2$
  boundaries}

Invariance of the $\textnormal{dS}_2$ and $\textnormal{AdS}_2$ boundaries
under $\dl_r$ can also be checked using the same arguments as in
ref.~\cite{HS}. Let us briefly see this.

Consider first the case of $\textnormal{dS}_2$. The boundary is located in
Poincar\'e coordinates at \hbox{$t=0$}, equivalently\, $x^+\!+x^-\!=0$. Since
the boundary must remain unchanged under conformal diffeomorphisms
$x^\pm\to x^\pm+\xi^\pm(x^\pm)$, the vector fields $\xi^\pm(t,x)$ must satisfy
\begin{equation}
  \pa_+^n\,\xi^+(0,x)= (-1)^{n+1}\pa_-^n\,\xi^-(0,x)\,,\qquad n=0,1,2,\ldots
  \label{diff-dS-restriction}
\end{equation}
One allows for field configurations of $\varphi$ and $a$ that behave near
$t=0$ as the $\textnormal{dS}_2$ vacuum solution of Section~3,
\begin{equation}
  \varphi,\,-\frac{R_0\,a}{2F_0\ell^2} =\ln\Big(\frac{\ell}{x^+\!+x^-}\Big)
  + O(1)\,,
    \label{dS-varphi-a-boundary}
\end{equation}
which satisfy the $\textnormal{dS}_2$ boundary condition~(\ref{dS-C}),
\begin{equation}
  0 = - (\om_- + \om_+) + \frac{1}{2\ell^2}\,(A_- + A_+)\bigg\vert_{t=0}
  = \pa_x \Big( \varphi - \frac{a}{2\ell^2}\Big)\bigg\vert_{t=0}\,.
  \label{dS-back}
\end{equation}
We must check that eq.~(\ref{dS-back}) is invariant under $\dl_r$. To do this,
compute first
$\dl_r(\om_{+\!}+\om_-)\big\vert_{t=0}$. Equation~(\ref{delta-omega-res})
gives for $\dl_r(\om_{+\!}+\om_-)$ two contributions, one from the Lie
derivative ${\cal L}_{\xi_r}(\om_{+\!}+\om_-)$, and one from the boost
generated terms $- \frac{1}{2}\,\big(\pa_+^2\xi^+ -
\pa_-^2\xi^-\big)$. Expanding in powers of $t$, noting that near $t=0$
\begin{equation}
  \pa^n_\pm\xi^\pm(t\pm\/x) =   \pa^n_\pm \xi^\pm(0,x)
  +  t\, \pa^{n+1}_\pm \xi^\pm(0,x)+ O(t^2)\,,
\end{equation}
and recalling eq.~(\ref{diff-dS-restriction}), it is very easy to see that the
Lie derivative takes at the boundary the value
\begin{equation}
  {\cal L}_{\xi_r}(\om_{+\!}+\om_-)\,\bigg\vert_{t=0}
  =\frac{1}{2}\,\big[\, \pa_+^2\xi^+(0,x) - \pa_-^2\xi^-(0,x) \,\big]\,.
  \label{failed-one}
\end{equation}
This cancels the contribution from boosts and gives
\begin{equation}
  \dl_r(\om_{+\!}+\om_-)\bigg\vert_{t=0} =
  {\cal L}_{\xi_r}(\om_++\om_-) 
  - \frac{1}{2}\,\pa_+^2\xi^+ 
  + \frac{1}{2}\,\pa_-^2\xi^-\bigg\vert_{t=0} = 0\,.
  \label{failed-two}
\end{equation}
Analogous arguments show that $\dl_r(A_{+\!}+A_-)\big\vert_{t=0} =0$, thus
completing the proof of invariance of condition~(\ref{dS-back}) under
$\dl_r$.

The proof for an $\textnormal{AdS}_2$ boundary goes along the same lines. The
only differences are that now the boundary is at $x=0$, equivalently
$x^{+\!}-x^-=0$, eq.~(\ref{diff-dS-restriction}) is replaced with~\cite{HS}
\begin{equation}
  \pa_+^n\,\xi^+(t,0)= \pa_-^n\,\xi^-(t,0)\,,\qquad n=0,1,2,\ldots,
  \label{diff-AdS-restriction}
\end{equation}
and the boundary condition takes the form~(\ref{AdS-C}). Taking into account
these changes, and proceeding as for $\textnormal{dS}_2$ one has\,
\hbox{$\dl_r(\om_{-\!}-\om_+)\big\vert_{x=0}=0$}\,and\,
$\dl_r(A_{-\!}-A_+)\big\vert_{x=0}=0$.

\subsection{Conserved currents, charges, and Hamiltonian formalism}

The field equations that result upon taking variations
with respect to $\varphi$ and $a$ in the action~(\ref{action-conformal}) are
\begin{align}
  \pa_+\pa_{-\!}\ast{\!F} -\frac{1}{4}\,e^{2\varphi}\,\Big[
  R\ast{\!F} +\frac{1}{2\ell^2}\,(\ast{F})^2- \frac{\gamma}{\ell^2}\Big]
  +\, \kappa\, T^{\,m}_{+-}
  &=0\,, \label{field-eq-varphi}\\[3pt]
  \frac{1}{\kappa}\,\pa_+\pa_{-\!} \Big(R+\frac{\ast{F}}{\ell^2}\Big)
  + \pa_+J^m_{-}-\pa_-J^m_{+} &=0\,. \label{field-eq-a}
\end{align}
Equation~(\ref{field-eq-a}) can be written in terms of the  total $U(1)$ current
\begin{equation}
  J^{\scriptscriptstyle T}_\pm = \mp \frac{1}{2\kappa}\,\pa_\pm
      \Big(R+\frac{\ast{F}}{\ell^2}\Big) 
     + J^m_\pm 
  \label{J-total-conformal}
\end{equation} 
given by eq.~(\ref{J-total-mu}) as
$\pa_-J^{\scriptscriptstyle T}_+=\pa_+J^{\scriptscriptstyle T}_-$. This and
the conservation equation~(\ref{conservation-J}), which in the conformal gauge
reads $\pa_+J^m_-+\pa_-J^m_+=0$, imply
\begin{equation}
   \pa_-J^{\scriptscriptstyle T}_+=\pa_+J^{\scriptscriptstyle T}_-=0\,.
\label{J-holorphic-conservation}
\end{equation} 

Consider the case of no additional matter. Standard methods show that the
Noether currents preserved by $\dl_r^\pm$ are
\begin{equation}
  \tilde{T}^g_{\pm\pm} = - \frac{1}{\kappa}
  \bigg[ \pa_\pm\pa_{\pm\!}\ast{\!F}
    -  2\,\pa_\pm\varphi\;\pa_{\pm\!}\ast{\!F}
    - \frac{F_0\ell^2}{R_0}\,\pa_\pm\pa_\pm\Big(R+\frac{\ast{F}}{\ell^2}\Big)
    +  \,\pa_\pm\/a\;\pa_\pm\Big(R+\frac{\ast{F}}{\ell^2}\Big)\bigg]\,.
    \label{improved-T}
\end{equation}
In fact, using eqs.~(\ref{field-eq-varphi}) and~(\ref{field-eq-a}), it is
straightforward to see that
\begin{equation}
 \pa_-\tilde{T}^g_{++}=\pa_+\tilde{T}^g_{--}=0\,.
  \label{holomorphic-conservation}
\end{equation}
The currents $\tilde{T}^g_{\pm\pm}$ can also be cast as
\begin{equation}
  \tilde{T}^g_{\pm\pm} = {T}^g_{\pm\pm}
  \pm \frac{2F_0\ell^2}{R_0}\,\pa_\pm J^g_\pm \pm 2\,J^g_\pm\,\pa_\pm a,
  \label{sum-improved-T}
\end{equation}
where $T^g_{\pm\pm}$ are obtained from eq.~(\ref{sin-2}), and
\begin{equation}
  J^g_\pm=\mp \frac{1}{2\kappa}\,\pa_\pm
      \Big(R+\frac{\ast{F}}{\ell^2}\Big)
\end{equation}
are the gravity contributions to the $U(1)$ current.
The corresponding conserved charges are
\begin{equation}
  Q^\pm = \int\! dx^\pm\,  T_{\pm\pm}(x^+)\; \xi^\pm(x^+) \,.
  \label{charges}
\end{equation}
Let us check that $Q^\pm$ generate through Poisson brackets residual
transformations,
\begin{equation}
  \dl_r^\pm \phi=\big\{Q^\pm,\phi\big\}\,, \qquad \phi=\varphi,a.
  \label{generator}
\end{equation}

The action $S_{\scriptscriptstyle CFT}$ can be regarded as describing a
dynamical system with Lagrangian
\begin{equation}
 S_{\scriptscriptstyle CFT}=\int\!dt\,L\,,\qquad
   L=2\!\int \!dx\, {\cal L}_{\scriptscriptstyle CFT}\,,
   \label{Lagrangian}
\end{equation}  
where ${\cal L}_{\scriptscriptstyle CFT}$ is the integrand in
eq.~(\ref{action-conformal}). Since the Lagrangian $L$ contains second
derivatives with respect to time of $\varphi$ and $a$, the Hamiltonian
formulation is a bit more involved than for dynamical systems with only
first-order time derivatives; see e.\,g. refs.~\cite{Morozov,Govaerts} for
reviews. The phase space is now formed by the generalized coordinates
\begin{equation}
  q^\varphi_0=\varphi(t,x),~~ q^a_0=a(t,x),~~
  q^a_1=\dot{a}(t,x),~~ q^\varphi_1=\dot{\varphi}(t,x)
\end{equation}
and their conjugate momenta, 
\begin{equation}
  \pi_\phi^0(t,x)= \frac{\pa L}{\pa \dot{q}^\phi_0(t,x)}
  -\pa_t\,\frac{\pa L}{\pa \dot{q}^\phi_1(t,x)}\,,\qquad
  \pi_\phi^1(t,x)=\frac{\pa L}{\pa \dot{q}^\phi_1(t,x)}\,,
\end{equation}
where we have introduced the index $\phi=\varphi,a$. The Poisson brackets are
the usual ones 
\begin{gather}
  \big\{ q^\phi_{\,i}(x,t),\, \pi^j_{\phi'}(y,t)\big\} =
  \dl^\phi_{\,\phi'}\,\dl^j_{\,i}\,\dl(x-y)\,, \label{nontrivial-PB}\\
 \big\{ q^\phi_{\,i}(x,t),\, q^{\phi'}_{\,j}(y,t)\big\} =
\big\{ \pi^i_\phi(x,t),\, \pi^j_{\phi'}(y,t)\big\} =0\,,
  \label{trivial-PB}
\end{gather}
with\, $\phi,\phi'=\varphi,a$ \,and\, $i,j=0,1$.  And finally, the Hamiltonian
reads
\begin{equation}
  H = \int\! dx \left( \pi^0_\varphi \,\dot{q}^\varphi_0 + \pi^0_a\, \dot{q}^a_0
     + \pi^1_\varphi \,\dot{q}^\varphi_1 + \pi^1_a \,\dot{q}^a_1 \right) - L\,,
\end{equation}
and Hamilton's equations take the form
\begin{equation}
  \dot{q}^\phi_i = \frac{\pa H}{\pa \pi^i_\phi}\,,\qquad
  \dot{\pi}^i_\phi = -\,\frac{\pa H}{\pa q^\phi_i}\,,\qquad
  \phi=\varphi,\,a,\quad i=0,1.
  \label{Hamilton-equations}
\end{equation} 
Some simple calculations give
\begin{alignat}{6}
  \pi^0_\varphi&=-\frac{1}{\kappa}\,\pa_t\!\ast\!F,&&\qquad&&
  \pi_\varphi^1= \frac{1}{\kappa}\ast\!F ,\label{pi-0}\\[3pt]
  \pi^0_a&=\frac{1}{2\kappa}\,\pa_t
  \Big( R + \frac{\ast\/F}{\ell^2}\Big) ,&& &&
  \pi_a^1= -\frac{1}{2\kappa}\,
  \Big( R + \frac{\ast\/F}{\ell^2}\Big) \label{pi-1}
\end{alignat}
for the momenta, and  
\begin{equation}
  H   = \int\! dx \bigg[ \pi^0_\varphi\, q^\varphi_1 + \pi^0_a\, q^a_1
  + \pi^1_\varphi \, \pa_x^2q^\varphi_0  + \pi^1_a \, \pa_x^2q^a_0   
      - \kappa\,e^{2\varphi} \bigg(\pi^1_a
      + \frac{\pi^1_\varphi}{4\ell^2}\bigg)\pi^1_\varphi
      -\frac{\gamma}{2\kappa\ell^2}\,e^{2\varphi} \bigg]
\end{equation}
for the Hamiltonian.  It is straightforward to check that the Hamilton
equations reproduce the same field equations~(\ref{field-eq-varphi})
and~(\ref{field-eq-a}) as the variational approach. The Poisson brackets in
turn imply that
\begin{gather}
  \big\{ \varphi(t,x)\,,\,\pa_\pm\!\ast\!F(t,y)\big\} =
   - \big\{\pa_\pm\varphi(t,x)\,,\, \ast F(t,y)\big\} =
   -\frac{\kappa}{2}\,\dl(x-y), \label{PB-varphi-pa-F} \\[3pt]
   \Big\{ a(t,x)\,,\,\pa_\pm\Big( R + \frac{\ast F}{\ell^2}\Big)(t,y)\Big\}
   = \Big\{ \pa_\pm a(t,x)\,,\,
        \Big(R + \frac{\ast F}{\ell^2}\Big)(t,y) \Big\}\
   = \kappa\,\dl(x-y). \label{PB-a-pa-RF}
 \end{gather}
 Using these, one easily verifies that eqs.~(\ref{generator})
 hold. Furthermore, the currents $T_{\pm\pm}$ satisfy the the equal-time
 bracket
 \begin{equation}
   \big\{\, T_{++}(x^+)\,,\, T_{++}(y^+)\big\} =
    \frac{1}{\kappa}\,\dl(x^+-y^+) \,\pa_+ T_{++}(x^+)
         + \frac{2}{\kappa}\,\,T_{++}(x^+)\,\pa_+\dl(x^+-y^+) \,,
\label{current-algebra}
\end{equation}
and a similar expression for $T_{--}$. This is analogous to JT-Maxwell
gravity~\cite{HS}.

\subsection{Matter and central charge in the quantum theory}

The argument for the occurrence of a central charge in JT-Maxwell
gravity~\cite{HS} also holds for our model. Let us briefly go through it. If
matter is included, instead of $\tilde{T}^g_{\pm\pm}$ in
eq.~(\ref{sum-improved-T}), one has
\begin{equation}
  \tilde{T}_{\pm\pm} = {T}^g_{\pm\pm} + {T}^m_{\pm\pm}
  \pm \frac{2F_0\ell^2}{R_0}\,\pa_\pm J^{\scriptscriptstyle T}_\pm
  \pm 2\,J^{\scriptscriptstyle \scriptscriptstyle T}_\pm\,\pa_\pm a.
  \label{sum-improved-T-matter}
\end{equation}
For reasonable choices of matter, one expects the following:
\setlist{nolistsep}
\begin{enumerate}
\item[i)] ${T}_{\pm\pm\!} ={T}^g_{\pm\pm\!} + {T}^m_{\pm\pm}$ will be
  holomorphically conserved. Equation~(\ref{J-holorphic-conservation}) and the
  constraint~(\ref{J-total-mu}) then imply\, $\pa_\mp\tilde{T}_{\pm\pm\!}=0$.
\item[ii)] $S_m$ will have a contribution $\sqrt{|g|}J^mA$. This produces a
  term\, $\mp 2J^m_\pm\,\pa_\pm a$ \,in\, ${T}^m_{\pm\pm}$ \,that cancels the
  contribution\, $\pm 2J^m_\pm\,\pa_\pm a$ \,hidden in the fourth term in
  eq.~(\ref{sum-improved-T-matter}).
\end{enumerate}
All things together, the conserved matter current\footnote{As implied by
  eqs.~(\ref{J-holorphic-conservation}) and~(\ref{conservation-J})}
$\pa_-J^m_+=\pa_+J^m_-=0$ \,enters $\tilde{T}_{\pm\pm}$ through\,
$\pm \pa_\pm J^{m}_\pm$ \,with coefficient \,${2F_0\ell^2}/{R_0}$. Assume now,
as in ref.~\cite{HS}, that the current is anomalous so that in the quantum
theory
\begin{equation}
   \big[ J^m_+(x^+)\,,\,J^m_+(y^+)\big] = -k\pa_+\,\dl(x^+-y^+).
\end{equation}
The current algebra~(\ref{current-algebra}) will then have a central term 
\begin{equation}
  F_0^2\ell^4k\,\pa_+^3\,\dl(x^+-y^+),
\end{equation}
where we have used that $R^2_0=4$ for our choice of Poincar\'e coordinates.
The result is formally the same for $\textnormal{dS}_2$ and
$\textnormal{AdS}_2$ backgrounds, but it remains to find explicit
realizations.

\section{Further remarks and conclusions}

\subsection{The Euclidean case}

The same model can be formulated with Euclidean signature. The starting point
for the Utiyama-Kibble-Sciama procedure is now the central extension
$\mathfrak{e}_{0}={Span}\{P_1, P_2,J,Q\}$ of the Euclidean algebra in two
dimensions, or Nappi-Witten algebra~\cite{Nappi-Witten}, whose Lie bracket is
\begin{equation}
  [P_1,P_2]=Q\,,  \qquad [J,P_1]=P_2\,,
  \qquad [J,P_2]=-P_1\,,  \qquad[Q,P_a]=[Q,J]=0\,.
  \label{NW-algebra}
\end{equation}
The classical action is the same as in eq.~(\ref{action-intro}), except for
the sign in front of $F^2$, which is now positive since the right-hand side of
eq.~(\ref{identity-epsilon}) changes its sign for Euclidean signature. Vacuum
solutions are either black hole type or have constant scalar curvature and
constant $\ast\/ F$, in which case they are locally isomorphic to 2d Euclidean
space, the sphere or the hyperbolic plane.

\subsection{No-go results for other 2d Yang-Mills gravity models}

Powers of $R$ and/or $\ast F$ can be included in the action $S$ in
eq.~(\ref{action}) without changing the symmetry of the model. The question
arises as to whether there are models invariant under
\hbox{${\dl}_{(\xi,\Sig)}={\cal L}_\xi+\tilde{\dl}_\Sigma$}, with $\Sigma$
taking values in the two-dimensional non-Abelian algebra
$\mathfrak{na}_2$\footnote{Up to isomorphisms, there is only one
  two-dimensional non-Abelian real Lie algebra, namely $[X,Y]=Y$.}. In this
case the closure relation would no longer be~(\ref{closure}) but rather
\begin{equation}
  \big[\,{\cal L}_{\xi_1}\! +  \tilde{\dl}_{\Sig_1}\,,
       \,{\cal L}_{\xi_2} \!+ \tilde{\dl}_{\Sig_2}\,]
  = {\cal L}_{[\xi_1,\xi_2]} + 
    \tilde{\dl}_{[\Sig_1,\Sig_2] + {\cal L}_{\xi_2}\Sig_1-{\cal L}_{\xi_1}\Sig_2}\,.
  \label{semidirect}
\end{equation}
In the sequel we provide an answer to this question in the negative. We show
in particular that there is no real four-dimensional Lie algebra whose gauging
as described in Section~2 leads to an invariant action linear in the Riemann
curvature.

The proof is by inspection. We are interested in indecomposable
four-dimensional real Lie algebras that have a non-Abelian two-dimensional
algebra $\mathfrak{na}_2$ as a subalgebra. All such algebras are solvable and
are listed in the literature see e.\,g.~ref~\cite{Andrada}. Some care must be
taken though, since some of them have more than one $\mathfrak{na}_2$
subalgebra and different choices for $\mathfrak{na}_2$ lead to different
semidirect products ${\cal X}\!\ltimes\! \mathfrak{na}_2$. Let us illustrate
this with an example. Consider the Lie algebra
${Span}\hspace{0.2pt}\{t_0, t_1, t_2, t_3\}$, with
\begin{equation}
  \mathfrak{p}_\lambda\!:~~
  [t_0, t_1] = \lambda t_1,~~ [t_0, t_2] = (1 - \lambda)\,t_2, ~~ 
     [t_0, t_3] = t_3, ~~ [t_1, t_2] = t_3,  ~~ \lambda \geq\frac{1}{2}\,. 
     \label{11-deformed}
\end{equation}
Note that for $\la=1$ and $t_0=J$, $t_1=P_0$, $t_2=Q$ and $t_3=P_1$ the
central extension of the 2d Poincar\'e algebra in eqs.~(\ref{11-algebra}) is
recovered. Substituting 
\begin{align}
  B_\mu&=b^0{\!}_\mu\, t_0  + b^1{\!}_\mu\, t_1 +  b^2{\!}_\mu\, t_0
         +  b^3{\!}_\mu \, t_3\,,
  \label{A-form-def} \\[3pt]
  G_{\mu\nu}&=G^0{\!}_{\mu\nu}\, t_0  + G^1{\!}_{\mu\nu}\, t_1
      + G^ 2{\!}_{\mu\nu}\, t_2 +G^3{\!}_{\mu\nu}\, t_3
  \label{G-form-def}
\end{align}
in $G_{\mu\nu}=\pa_\mu B_\nu-\pa_\nu B_\mu +[B_\mu,B_\nu]$, we have
\begin{align}
  G^0{\!}_{\mu\nu} &= \pa_\mu\/b^0{\!}_\nu - \pa_\nu\/b^0{\!}_\mu\,  \label{G0}\\[2pt]
  G^1{\!}_{\mu\nu} &= \pa_\mu\/b^1{\!}_\nu - \pa_\nu\/b^1{\!}_\mu
    + \lambda \, ( b^0{\!}_\mu\/b^1{\!}_\nu -b^0{\!}_\nu\/b^1{\!}_\mu) \,
    \label{G1}\\[2pt]
    G^2{\!}_{\mu\nu}  &= \pa_\mu\/b^2{\!}_\nu - \pa_\nu\/b^2{\!}_\mu
    = (1-\lambda)\,( b^0{\!}_\mu\/b^2{\!}_\nu -b^0{\!}_\nu\/b^2{\!}_\mu) \,,
    \label{G2}\\[2pt]
    G^3{\!}_{\mu\nu}  &= \pa_\mu\/A_\nu -  \pa_\nu\/A_\mu
   + ( b^0{\!}_\mu\/b^3{\!}_\nu -b^0{\!}_\nu\/b^3{\!}_\mu)
   + ( b^1{\!}_\mu\/b^2{\!}_\nu -b^1{\!}_\nu\/b^2{\!}_\mu) \,.
   \label{G3}
\end{align}
There are three possible choices for the two-dimensional non-Abelian subalgebra
$\mathfrak{na}_2$,
\begin{equation}
  \textnormal{(a)}~[t_0, t_3] = t_3, \qquad
  \textnormal{(b)}~[t_0, t_1] = \lambda t_1,\qquad
  \textnormal{(c)}~[t_0, t_2] = (1 - \lambda)\,t_2.
  \label{5-10}
\end{equation}
Making $\lambda\to 1-\lambda$, $t_1\to -t_2$ and $t_2\to t_1$, the commutator
(\ref{5-10}c) reduces to (\ref{5-10}b) while keeping all other commutators in
eq.~(\ref{11-deformed}) unchanged. Hence it is enough to consider cases
(\ref{5-10}a) and (\ref{5-10}b).

(a) Case $\mathfrak{na}_2=Span\hspace{0.1pt}\{t_0,t_3\}$. Under
$\dl_{(\xi,\Sig)}$, with $\Sig= \th t_0 + \tau t_3$, the gauge fields transform as
\begin{align}
  \dl\/b^0{\!}_{\mu} &= {\cal L}_\xi b^0{\!}_{\mu} + \pa_\mu\th \,,  \label{G0-var}\\[2pt]
  \dl\/b^1{\!}_{\mu} &= {\cal L}_\xi b^1{\!}_{\mu} 
    - \lambda \, b^1{\!}_{\mu}\, \th\, ,   \label{G1-var} \\[2pt]
  \dl\/b^2{\!}_{\mu}  &= {\cal L}_\xi b^2{\!}_{\mu} 
     -(1-\lambda) \,b^2{\!}_{\mu}\, \th \,, \label{G2-var}\\[2pt]
    \dl\/b^3{\!}_{\mu}  &= {\cal L}_\xi b^3{\!}_{\mu} + \pa_\mu\tau 
   - b^3{\!}_{\mu} \,\th +  b^0{\!}_{\mu}\, \tau\,,  \label{G3-var}
\end{align}
whereas the variations of the field strengths read
\begin{align}
  \dl\/G^0{\!}_{\mu\nu} &= {\cal L}_\xi G^0{\!}_{\mu\nu} \,,  \label{G0-var}\\[2pt]
  \dl\/G^1{\!}_{\mu\nu} &= {\cal L}_\xi G^1{\!}_{\mu\nu} 
    - \lambda \, G^1{\!}_{\mu\nu}\, \th\, ,   \label{G1-var} \\[2pt]
  \dl\/G^2{\!}_{\mu\nu}  &= {\cal L}_\xi G^2{\!}_{\mu\nu}
     -(1-\lambda) \,G^2{\!}_{\mu\nu}\, \th \,, \label{G2-var}\\[2pt]
    \dl\/G^3{\!}_{\mu\nu}  &= {\cal L}_\xi G^3{\!}_{\mu\nu} 
   - G^3{\!}_{\mu\nu} \,\th +  G^0{\!}_{\mu\nu}\, \tau\,.  \label{G3-var}
\end{align} 
For $\lambda\neq 1$, the only invariants up to order 2 in the field
strengths are $\ast G^0$ and $(\ast G^0)^2$. The first one is a total
derivative that we ignore, while the second one gives a free theory for
$b^0{\!}_\mu$. A zweibein postulate that linearly maps $b^0{\!}_\mu$ to an
affine connection, $G^0{\!}_{\mu\nu}$ to the Riemann tensor and
$(G^1{\!}_{\mu\nu},G^2{\!}_{\mu\nu})$ to the torsion does exist. However,
since there is no nonfree invariant action, it will not lead to a 2d gravity
model.

(b) Case $\mathfrak{na}_2=Span\hspace{0.1pt}\{t_0,t_3\}$. Taking now
$\Sig= \th t_0 + \tau t_1$, the transformation laws are
\begin{align}
  \dl\/b^0{\!}_{\mu} &= {\cal L}_\xi b^0{\!}_{\mu} + \pa_\mu\th \,,  \label{G0-var}\\[2pt]
  \dl\/b^1{\!}_{\mu} &= {\cal L}_\xi b^1{\!}_{\mu} + \pa_\mu\tau 
    + \lambda \, (b^0{\!}_{\mu}\, \tau - b^1{\!}_{\mu}\, \tau),    \label{G1-var} \\[2pt]
  \dl\/b^2{\!}_{\mu}  &= {\cal L}_\xi b^2{\!}_{\mu} 
     -(1-\lambda) \,b^2{\!}_{\mu}\, \th \,, \label{G2-var}\\[2pt]
    \dl\/b^3{\!}_{\mu}  &= {\cal L}_\xi b^3{\!}_{\mu}  
   - b^3{\!}_{\mu} \,\th - b^2{\!}_{\mu}\, \tau\,,  \label{G3-var}
\end{align}
and 
\begin{align}
  \dl\/G^0{\!}_{\mu\nu} &= {\cal L}_\xi G^0{\!}_{\mu\nu} \,,  \label{G0-var}\\[2pt]
  \dl\/G^1{\!}_{\mu\nu} &= {\cal L}_\xi G^1{\!}_{\mu\nu} 
    + \lambda \, (G^0{\!}_{\mu\nu}\, \tau - G^1{\!}_{\mu\nu}\, \tau),  \label{G1-var} \\[2pt]
  \dl\/G^2{\!}_{\mu\nu}  &= {\cal L}_\xi G^2{\!}_{\mu\nu}
     -(1-\lambda) \,G^2{\!}_{\mu\nu}\, \th \,, \label{G2-var}\\[2pt]
    \dl\/G^3{\!}_{\mu\nu}  &= {\cal L}_\xi G^3{\!}_{\mu\nu} 
   - G^3{\!}_{\mu\nu} \,\th - G^2{\!}_{\mu\nu}\, \tau\,.  \label{G3-var}
\end{align} 
It is clear from this last set of  equation that the same conclusion as in
case (a) holds. 

Going through the list of solvable four-dimensional real Lie
algebras~\cite{Andrada}, we have found that the only invariants that occur are
either a total derivative or provide a free theory for a $B_\mu$
component. All this speaks in favor of the uniqueness of the model in
Section~2 within the class of Yang-Mills type models for 2d gravity.

\section*{Acknowledgments}
The authors wish to thank Thomas Hartman for correspondence. This work was
partially funded by the Spanish Ministry of Education and Science through
grant PGC2018-095382-B-I00. S.\,A. acknowledges
Universidad Complutense Madrid and Banco Santander for
support through a predoctoral fellowship.

\newlength{\bibitemsep}\setlength{\bibitemsep}{.3\baselineskip plus
  .05\baselineskip minus .05\baselineskip}
\newlength{\bibparskip}\setlength{\bibparskip}{0pt}
\let\oldthebibliography\thebibliography \renewcommand\thebibliography[1]{%
  \oldthebibliography{#1}%
  \setlength{\parskip}{\bibitemsep}%
  \setlength{\itemsep}{\bibparskip}%
}

\end{document}